# An Interactive Google Earth Engine Application for Global Multi-Scale Vegetation Analysis Using NDVI Thresholding

Md. Moktader Moula[*a], Israt Jahan Shonom[a], Azharul Islam[b], Mohammad Mosharraf Hossain[a]

[a] Institute of Forestry and Environmental Sciences, University of Chittagong, Chittagong-4331, Bangladesh

[b] Strategic Landscape Planning and Management, School of Life Sciences, Weihenstephan, Technische Universität München, Emil-Ramann-Str. 6, 85354 Freising, Germany

**Corresponding author's email address**
moktader.ifes@std.cu.ac.bd

**Abstract**

Monitoring vegetation dynamics is crucial for addressing global environmental challenges like degradation and deforestation, but traditional remote sensing methods are often complex and resource-intensive. To overcome these barriers, we developed an interactive, cloud-based application on the Google Earth Engine (GEE) platform for few clicks on-demand global vegetation analysis without complex technical knowledge. The application automates the calculation of vegetated areas using the Normalized Difference Vegetation Index (NDVI) derived from Sentinel-2 and Landsat imagery. It utilizes a median composite of images over a selected period to create a single, robust, cloud-free image, minimizing atmospheric noise and other artifacts. It offers a flexible, global multi-scale analytical platform, allowing users to define regions of interest based on administrative boundaries, protected areas, or custom-drawn polygons. The user-friendly interface enables the selection of specific time periods and NDVI thresholds to quantify vegetation cover in real time, eliminating the need for manual and time intensive data handling and processing. A validation of the platform was conducted for two protected areas in Bangladesh which demonstrated high accuracy, with area estimates showing over 97% agreement with published reference data. By simplifying access to powerful geospatial analytics to general people, this tool provides a scalable and practical solution for researchers, land managers, policymakers, and any interested person to monitor vegetation trends, support conservation efforts, to inform decision making in spatial context where policy maker need to use insights in few clicks and inform environmental policy.

- Developed an interactive, cloud-based application on Google Earth Engine for on-demand global vegetation analysis.
- Automated the calculation of vegetated areas using NDVI derived from Sentinel-2 and Landsat imagery.
- Enabled users to define regions of interest, select time periods, and apply NDVI thresholds for real-time vegetation cover assessment.

# Graphical abstract

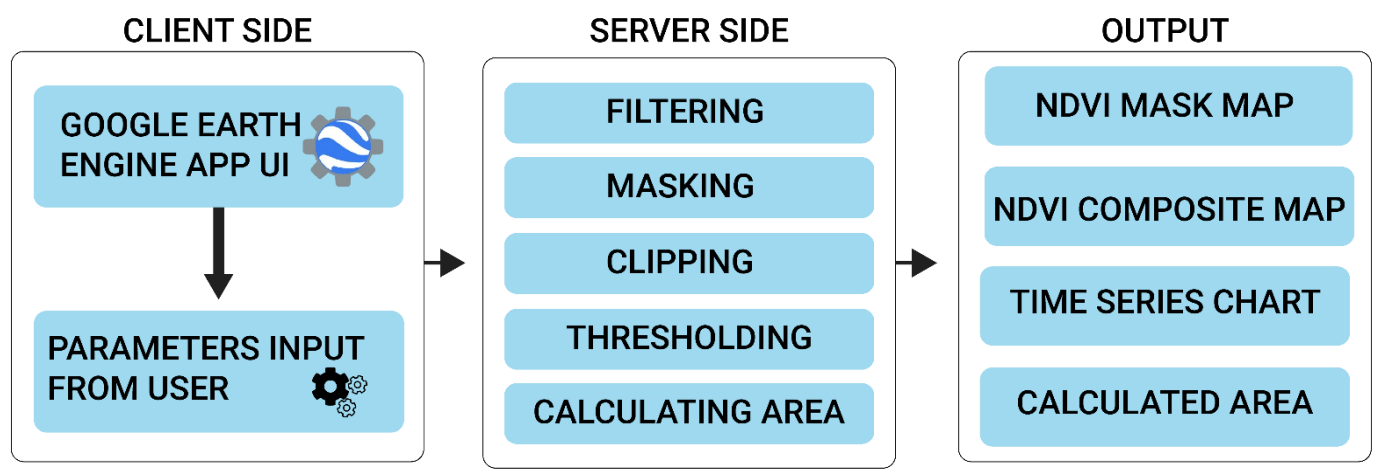

# Specifications table

| Subject area | Environmental Science |
|---|---|
| More specific subject area | *Remote Sensing for Vegetation Monitoring, Geospatial Analysis, and Environmental Conservation* |
| Name of your method | *Cloud-Based NDVI Vegetation Monitoring Tool on Google Earth Engine* |
| Name and reference of original method | *Tucker, C. J. (1979). Red and photographic infrared linear combinations for monitoring vegetation. Remote Sensing of Environment, 8(2), 127-150. (https://www.sciencedirect.com/science/article/abs/pii/0034425779900130)* |
| Resource availability | ***Software & Platform***<br>*Custom-built cloud-based application on Google Earth Engine (GEE) platform. Access available at:*<br>*https://ramizmoktader.users.earthengine.app/view/ndvibasedareaforestcoverbeta2*<br><br>***Data***<br>*Sentinel-2 imagery: Available on GEE(https://developers.google.com/earth-engine/datasets/catalog/COPERNICUS_S2_SR_HARMONIZED),*<br>*Landsat imagery: Available on GE (Ehttps://developers.google.com/earth-engine/datasets/catalog/landsat) ,*<br>*WPDA Protected Area Sources (PA): Available on GEE(https://developers.google.com/earth-engine/datasets/catalog/WCMC_WDPA_current_polygons),*<br>*GAUL dataset: Available on GEE(https://developers.google.com/earth-engine/datasets/tags/gaul),*<br>*Admin boundary of Bangladesh (level 0-4) (https://data.humdata.org/dataset/cod-ab-bgd)* |

## 1   Background

Forest degradation and deforestation are significant environmental issues, with an estimated 10 million hectares of forest lost annually (Ritchie, 2021). This loss poses serious threats to biodiversity, carbon storage, and critical ecosystem services, highlighting the need for accurate and timely monitoring of vegetation dynamics (Brockerhoff et al., 2017). Understanding these changes across various spatial and temporal scales is fundamental for effective environmental management, policy formulation, and sustainable development (Rees et al., 2001), and plays a critical role in a wide array of environmental applications, including forest conservation, agricultural productivity, drought monitoring, land use planning, and climate resilience (Franquesa et al., 2025). However, traditional methods for vegetation analysis are often computationally intensive, require specialized expertise, and are limited by data accessibility and processing complexity, creating a barrier for many potential users (Xue & Su, 2017).

To address these limitations, we developed an interactive application on the Google Earth Engine (GEE) platform that calculates vegetation area using Normalized Difference Vegetation Index (NDVI) thresholding. NDVI, derived from multispectral satellite imagery, is a widely accepted and one of the most commonly used indicators for assessing vegetation health and extent (Karnieli et al., 2010; Mutanga & Kumar, 2019; Pettorelli et al., 2005a). The tool uses GEE's cloud architecture and its multi-petabyte catalog of satellite imagery, which removes the need for users to manually download and process data (Gorelick et al., 2017). This approach directly addresses the major bottleneck of manual data handling and makes detailed vegetation analysis more accessible to people without extensive geospatial experience, such as policymakers, land managers, and researchers.

The application allows users to quantify vegetation cover through several spatial analysis modes, providing flexibility for different use cases. It supports analysis based on national and sub-national boundaries using FAO GAUL (Global Administrative Unit Layers) data, which is useful for country-level reporting and policymaking. For conservation-focused work, users can analyze vegetation within Protected Areas (PAs) from a global dataset to help assess conservation effectiveness and monitor habitat integrity (Hoffmann, 2022). Finally, the tool allows for analysis across custom regions defined by the user, enabling targeted research and localized project monitoring.

The system includes a comprehensive suite of satellite imagery from Sentinel-2 and Landsat, allowing users to select a dataset that fits their specific resolution and time-period requirements. The user workflow is designed to be intuitive: users can interactively define a region and time frame, then adjust NDVI thresholds and cloud cover limits to quantify and visualize vegetation cover in real time.

This tool differs from many conventional applications by providing a dynamic, web-based interface for analysis at multiple scales, rather than being limited to static datasets or specific locations. By simplifying the technical requirements, it can be a useful tool in fields such as forestry, agriculture, drought monitoring, and resource management. In summary, the application is designed to make vegetation analysis more straightforward, offering a practical method for monitoring vegetation resources and helping to bridge the gap between complex geospatial data and actionable insights.

## 2   Method details

The methodology integrates both client-side user interactions and server-side geospatial processing capabilities within the GEE platform. As illustrated in Figure 1, the workflow begins on the client side, where users define analysis parameters such as spatial extent, satellite data source, temporal range, NDVI threshold values, and maximum cloud cover. Once inputs are specified, the system performs a series of automated server-side operations including image collection filtering, cloud masking, NDVI computation, and threshold-based classification. The final outputs include a masked NDVI raster representing vegetated areas and a time-series plot summarizing vegetation dynamics, both of which are available for download.

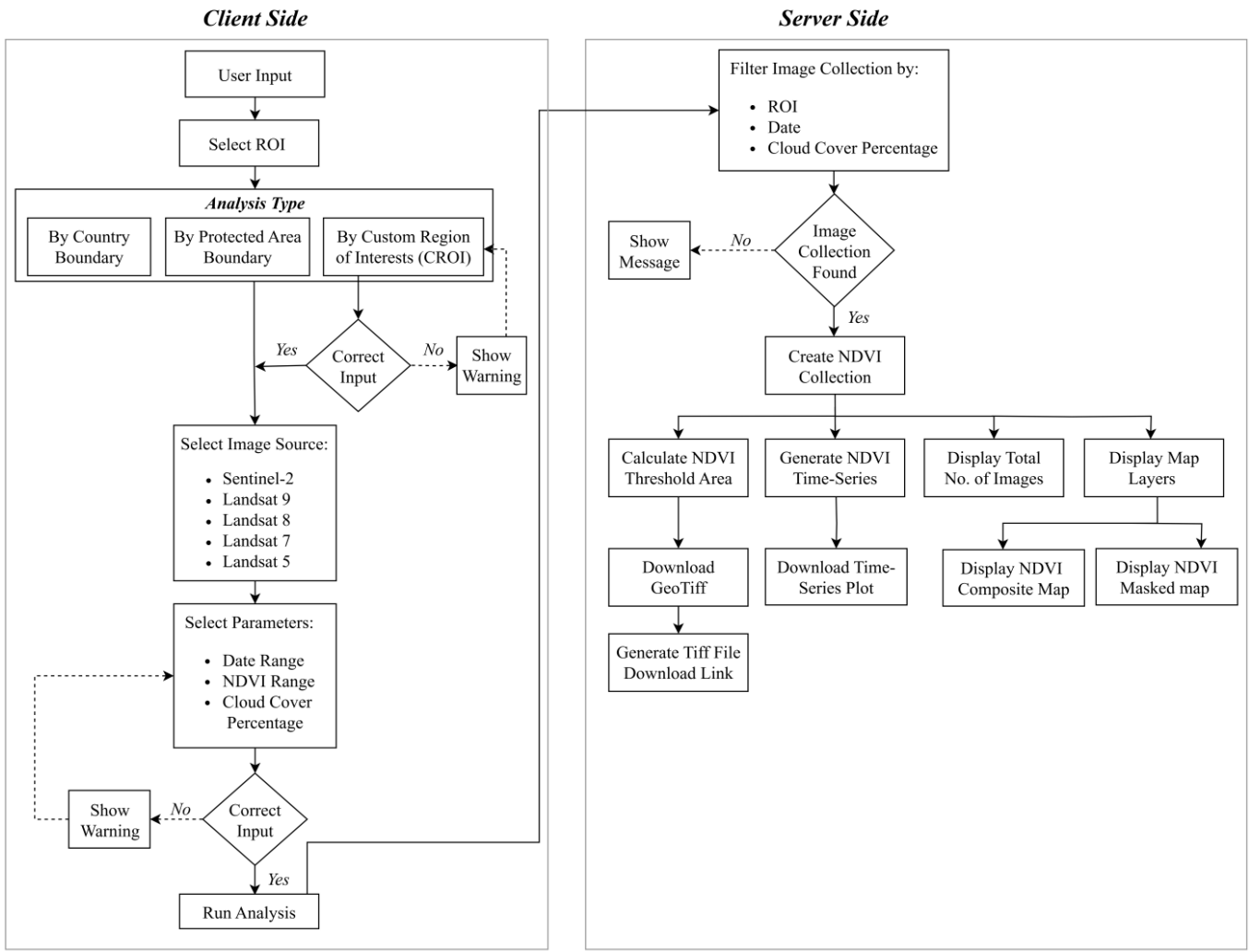

*Figure 1: Methodological flowchart of the study*

## 2.1 Delineation of the Region of Interest (ROI)

The analytical script facilitates the delineation of the geographical study area, or Region of Interest (ROI), through three distinct and flexible methods to ensure broad applicability.

### 2.1.1 Administrative Boundaries

ROIs can be defined using authoritative global and national vector datasets based on political and administrative units.

**Global Administrative Units**

For analyses at a global scale, the app utilizes the FAO Global Administrative Unit Layers (GAUL) 2015, specifically the Level 0 (countries) and Level 2 (sub-national) datasets (GEE ID: `FAO/GAUL/2015/level0` and `FAO/GAUL/2015/level2`, respectively). The application interface dynamically populates sub-national unit selections by filtering the GAUL collection based on the user's country selection and aggregating the distinct administrative names for the chosen sub-level (e.g., `ADM1_NAME`, `ADM2_NAME`).

**High-Resolution Boundaries for Bangladesh**

The application is designed for fine-scale analysis within Bangladesh, utilizing a hierarchical structure of administrative units. It enables users to precisely delineate an area of interest by navigating through four administrative levels:

Division, District, Upazila (Level 3), and Union (Level 4). To support this detailed analysis, the high-resolution boundaries for the Upazila and Union levels are sourced from custom shape files uploaded to the project's Google Earth Engine (GEE) assets. The interface implements a dynamic filtering system where selecting a unit at a higher level (e.g., a Division) populates the selection menu with the corresponding units at the next level down (e.g., its Districts), ensuring a seamless workflow to the most granular level.

### 2.1.2  Protected Areas

To support ecological and conservation research, ROIs can be defined using the World Database on Protected Areas (WDPA) (GEE ID: `WCMC/WDPA/current/polygons`). Protected areas are cornerstones of biodiversity conservation, crucial for safeguarding ecosystems, preserving genetic diversity, and maintaining essential ecological services (Dudley, 2008). This application's capacity for rapid, large-scale NDVI analysis presents a robust tool for monitoring the health and integrity of these vital areas. For instance, the time-series analysis can effectively monitor seasonal vegetation growth (phenology), detect the onset and severity of drought through negative NDVI anomalies, and identify long-term changes in forest cover that may indicate deforestation or degradation (Pettorelli et al., 2005b). The application first identifies the selected country's ISO 3166-1 alpha-3 code and then filters the WDPA dataset to retrieve a list of all designated protected areas within that nation's territory.

## 2.2  Custom User-Defined ROI (CROI)

To ensure maximum analytical flexibility, the application allows for the definition of custom ROIs through two mechanisms. Users can interactively delineate a polygon or rectangle directly onto the map interface. The resultant geometry is captured from the GEE `Map.drawingTools()` layer. Users can provide the GEE asset ID of a personal `ee.FeatureCollection()`. The app then ingests this asset and extracts its geometry to serve as the ROI.

## 2.3  Parameter Configuration and Validation

The integrity and efficiency of the analysis are contingent upon the rigorous configuration and validation of data and processing parameters.

### 2.3.1  Client-Side Input Validation

Prior to submitting computational tasks to the GEE servers, a suite of client-side validation checks is executed within the user's local environment. This pre-processing validation strategy significantly enhances the analytical workflow's efficiency, scalability, and robustness by preventing the submission of invalid parameters, thereby conserving computational resources and providing immediate feedback to the user. The validation checks include: the user-defined temporal range is validated against the operational period of the chosen satellite platform. For instance, a user cannot select a date prior to 2017-03-28 for a Sentinel-2 analysis.

The script verifies that the specified start date chronologically precedes or is the same as the end date. It also ensures that the user-defined minimum NDVI threshold is strictly less than the maximum NDVI threshold. The script confirms that numerical inputs are within their valid scientific ranges. It is critical to validate the NDVI range, as the index is mathematically constrained to values between -1 and +1. Values outside this range are theoretically impossible and indicate data or processing errors. Enforcing this constraint ensures the scientific validity of the analysis (Rouse et al., 1974). The application also validates that cloud cover is between 0 and 100 percent.

The application confirms that a Region of Interest has been successfully defined before allowing the analysis to proceed. Should any of these checks fail, the application displays an informative error message and visually highlights the problematic input field, guiding the user to correct the parameters without initiating a failed server-side task.

### 2.3.2  Satellite Data and Processing Parameters

The selection of appropriate satellite data and processing parameters is managed through a centralized configuration within the script.

The application supports analysis using imagery from five major medium to high-resolution satellite platforms. The script is configured with the specific GEE ImageCollection ID, native spatial resolution (scale), and relevant band names for each platform. To guide the user in selecting an appropriate temporal range, the application provides a summary of data availability for each sensor (Table 1).

*Table 1: Satellite Platform Data Availability*

| Satellite Platform | Start Date | End Date |
| --- | --- | --- |
| Sentinel-2 | 2017-03-28 | Present |
| Landsat 9 | 2021-10-31 | Present |
| Landsat 8 | 2013-04-11 | Present |
| Landsat 7 | 1999-05-28 | 2022-03-30 |
| Landsat 5 | 1984-03-16 | 2012-05-05 |

The user defines the temporal range (start and end dates), a maximum cloud cover percentage to filter the initial image collection, and minimum/maximum NDVI thresholds for the final classification.

### 2.4 Image Preprocessing and Cloud Masking

To ensure high data quality, rigorous preprocessing protocols are applied to the satellite imagery to remove pixels contaminated by clouds, cloud shadows, and other atmospheric artifacts.

#### 2.4.1 Initial Filtering

The designated `ee.ImageCollection()` is initially filtered based on the ROI's spatial bounds, the user-defined date range based on table 1, and the maximum cloud cover percentage using the collection's metadata property.

#### 2.4.2 Cloud Masking

A function is mapped over the filtered collection to apply a per-image mask based on the quality assessment (QA) band of each satellite.

**Sentinel-2**

For Sentinel-2 imagery, the Scene Classification Layer (SCL) band is used to generate a mask by retaining pixels classified as vegetation, non-vegetated, water, unclassified, and snow (classes 4, 5, 6, 7, 11), while discarding all others. Shadows, both from clouds and topography, are detected through SCL. By combining cloud and shadow masks with the SCL classification, only clear, illuminated surface pixels are retained, ensuring accurate vegetation indices, land cover classification, and remote sensing analysis.

**Landsat 8/9**

For Landsat 8 and 9 imageries, a bitwise operation is applied to the `QA_PIXEL` band to mask pixels flagged for dilated cloud, cirrus, cloud, and cloud shadow. Furthermore, surface reflectance values are derived by applying a scaling factor of $0.0000275$ and an offset of $-0.2$ to the raw digital numbers of the optical bands.

**Landsat 5/7**

A corresponding bitwise operation is applied to the `QA_PIXEL` band to mask cloud and cloud shadow. Furthermore, for these sensors, surface reflectance values are derived by applying a scaling factor of 0.0000275 and an offset of -0.2 to the raw digital numbers.

## 2.5 NDVI Calculation and Analysis

Subsequent to preprocessing, the core analytical procedures are executed.

### 2.5.1 Per-Image NDVI Calculation

The NDVI is based on the difference between the highest reflectance in the near infrared (NIR) spectral region due to leaf cellular structure and the maximum absorption of light in the visible red (R) due to chlorophyll pigments (Karnieli et al., 2010). For each cloud-masked image in the collection, NDVI is calculated via the `normalizedDifference()` function using the standard normalized difference formula shown aa Eq. 1 (Tucker, 1979),

$$NDVI = \frac{\rho_{NIR} - \rho_R}{\rho_{NIR} + \rho_R} \qquad Eq.\ 1$$

where $\rho$ is reflectance in the respective spectral bands. The specific Red and Near-Infrared (NIR) bands used for each sensor are detailed in Table 2.

*Table 2: Bands Used for NDVI Calculation*

| Satellite Platform | Red Band | NIR Band |
|---|---|---|
| Sentinel-2 | B4 | B8 |
| Landsat 9 & 8 | SR_B4 | SR_B5 |
| Landsat 7 & 5 | SR_B3 | SR_B4 |

### 2.5.2 Median Composite Image

To generate a single, robust, and spatially complete representation of the study period, a median composite is created. The `median()` reducer is applied to the collection of NDVI images, which calculates the median NDVI value at each pixel location across the time series. This method effectively mitigates noise from residual clouds, cloud shadows, and sensor artifacts that may remain after initial masking, as the median is a statistically robust measure, less sensitive to extreme outliers than the mean (Flood, 2013). This results in a cleaner, more representative image for analysis and classification.

### 2.5.3 NDVI Thresholding

The median NDVI composite is subsequently classified via thresholding, a standard image segmentation technique for delineating discrete land cover classes. This method is predicated on the principle that different surface types exhibit distinct NDVI value ranges. By defining specific thresholds, the application can automatically classify pixels into broad categories such as water, bare soil, or different densities of vegetation, allowing for rapid quantification of land cover types within the ROI. Table 3 provides a generalized guide to interpreting NDVI values for land cover classification.

*Table 3: Generalized NDVI Values for Land Cover Types (Source: Adapted from U.S. Geological Survey, 2018)*

| NDVI Range | Land Cover Type |
|---|---|
| ≤ 0.1 | Water, sand, snow, or barren |
| 0.2–0.5 | Grasslands, shrubs, sparse vegetation |
| 0.6–0.9 | Dense, healthy vegetation (e.g., forests) |

A final binary mask is generated wherein pixels with NDVI values greater than or equal to the user-defined minimum threshold and less than or equal to the maximum threshold are assigned a value of 1. The `selfMask()` function is then applied to retain only these pixels of interest for the final area calculation. This masked NDVI layer is used for computing the total vegetated area. In parallel, an NDVI composite image is generated by computing the median NDVI values across the filtered image collection, offering a visual representation of vegetation density and health.

## 2.6 Results, Visualization, and Export

The final stage of the workflow encompasses area quantification, trend visualization, and data product exportation.

### 2.6.1 Area Calculation

The total land area corresponding to the classified NDVI mask is quantified. This is achieved by multiplying the binary mask by `ee.Image.pixelArea()`, which generates an image where each pixel's value is its area. The `reduceRegion()` function, with an `ee.Reducer.sum()`, is then used to sum the area of all pixels within the ROI. The result is converted to square kilometers for reporting.

### 2.6.2 Data Export

The script provides functionality to export both the final binary NDVI mask layer and the NDVI composite layer as GeoTIFF raster. The `getDownloadURL()` method is invoked on-demand, generating a downloadable link with parameters for the export scale (matching the source satellite's resolution) and region (the defined ROI).

### 2.6.3 Time-Series Visualization

An interactive chart is generated using the `ui.Chart.image.series` function. The number of preprocessed images found are mentioned and the chart visualization plots the mean NDVI for the ROI against time for every image in the collection, allowing for the examination of vegetation phenology and long-term trends.

## 3 Method validation

To substantiate the analytical accuracy and operational performance of the developed application, a validation study was conducted. The study focused on two of Bangladesh's most prominent protected areas: Lawachara National Park, a critical biodiversity hotspot known for its primate populations, and Teknaf Wildlife Sanctuary, a key habitat for Asian elephants. These sites were strategically selected due to their ecological significance within the nation's protected area network and the availability of credible reference data for comparative analysis.

### 3.1 Validation Procedure and Data Processing

The validation procedure was initiated using the application's integrated module for selecting Regions of Interest (ROIs). This module sources its geospatial data from the World Database on Protected Areas (WDPA). The designated ROIs were processed through the application's full analytical workflow. For Lawachara National Park, the analysis utilized a composite of Sentinel-2 imagery from January 1 to March 31, 2021. For Teknaf Wildlife Sanctuary, the workflow processed Landsat 8 imagery from the same time frame, with a cloud cover threshold set to less than 10% (Figure 2). This automated process, as detailed in Sections 3 and 4, encompassed satellite imagery acquisition and filtering, application of cloud and shadow masking algorithms, calculation of the Normalized Difference Vegetation Index (NDVI), and subsequent land cover classification.

For the purpose of evaluating total land area coverage, a comprehensive classification was performed using the full theoretical NDVI range (–1.0 to +1.0). This approach ensures that all pixels within the ROI, regardless of surface type (e.g., dense vegetation, soil, water bodies), were included in the calculation. The total area for each ROI was computed within the Google Earth Engine (GEE) platform. The process utilized the `ee.Image.pixelArea()` function, which generates a raster layer where each pixel's value represents its precise geographical area. This area raster was then masked by the full-range NDVI composite to constrain the analysis strictly to the ROI's boundaries. Finally, the total area was quantified by aggregating the pixel values across the region using the `reduceRegion()` function with an `ee.Reducer.sum()` method. The resulting value, initially in square meters, was converted to square kilometers for reporting.

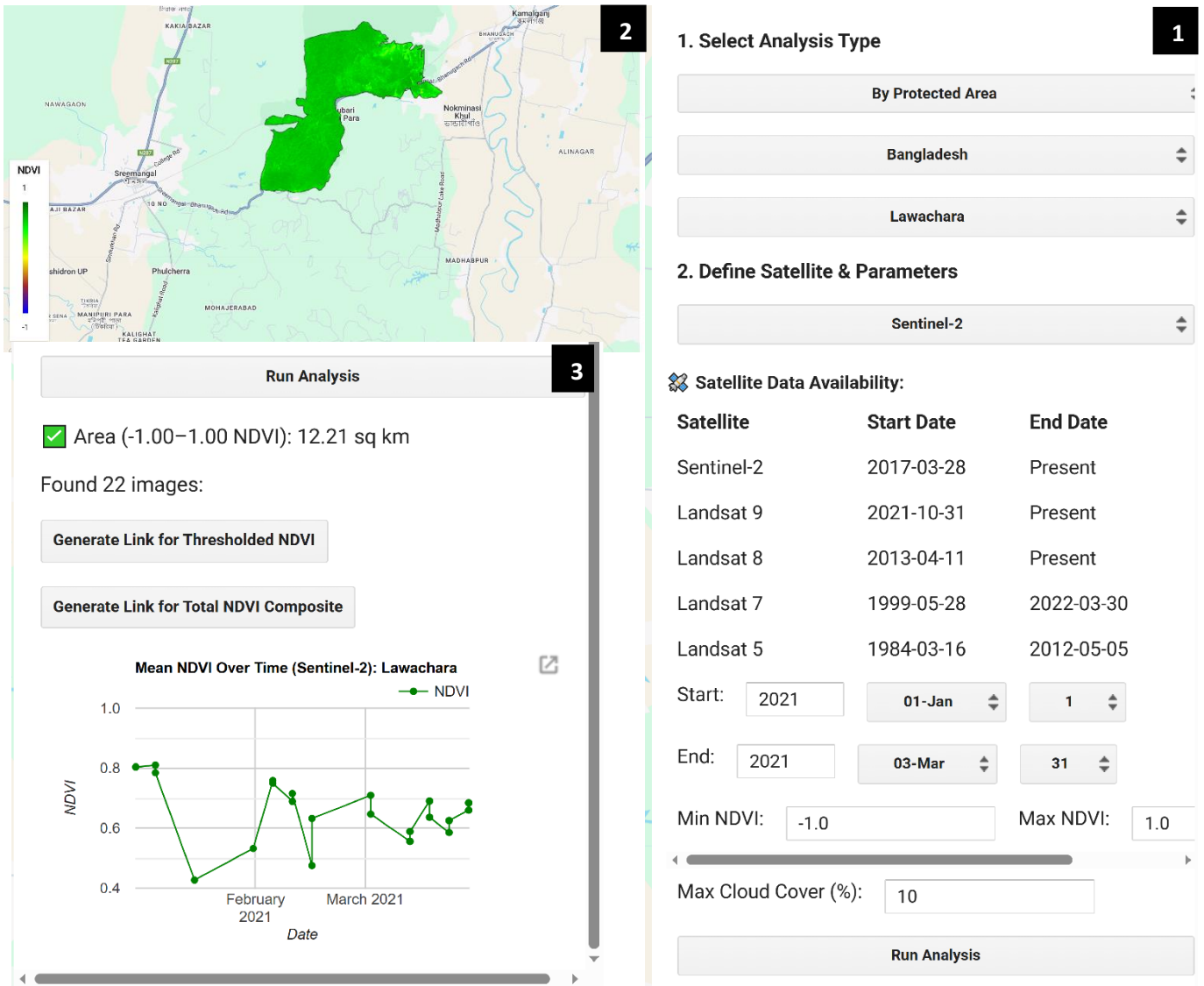

***Figure 2:*** *Workflow within GEE application to calculate area within NDVI range (–1.0 to +1.0) for Lawachara National Park of Bangladesh*

### 3.2   Results and Comparative Analysis

The total area estimates derived from the application's GEE-based workflow were systematically compared against officially recognized figures from peer-reviewed scientific literature. The results of this comparative analysis, presented in Table 4, demonstrate a high degree of congruence between the computed areas and the established reference values.

***Table 4:*** *Comparison of Application-Derived Area Estimates with Published Reference Data*

| Protected Area | Area calculated from Application (km²) | Reference Area (km²) | Source | Agreement (%) |
|---|---|---|---|---|
| **Lawachara National Park** | 12.21 | 12.50 | Islam et al. (2021) | 97.68 |
| **Teknaf Wildlife Sanctuary** | 117.53 | 116.15 | Ullah et al. (2022) | 98.80 |

The findings indicate that the application's area estimation methodology is highly consistent with authoritative figures, achieving agreement percentages exceeding 97%. The minor observed discrepancies (less than 2%) are well within acceptable limits for geospatial analysis and are attributable to several well-documented technical factors (Foody, 2002). Such variations are expected when comparing datasets from different sources and are likely due to subtle differences in boundary resolutions, the use of different map projection systems or geodetic datums, and the inherent sensitivity of NDVI-based classification at transitional land cover boundaries (e.g., forest edges).

## 4   Limitations

The application's performance is subject to the computational constraints of Google Earth Engine's free resources, which may cause it to lag during analyses of very large areas, such as an entire country. Furthermore, while the methodology is built on established and conventional equations, a lack of comparable studies in the existing literature made it difficult to validate the results on a large scale.

## 5   Conclusion

We developed an interactive Google Earth Engine application that simplifies and automates NDVI-based vegetation analysis. The tool provides a scalable, multi-scale solution for quantifying vegetation cover using Sentinel-2 and Landsat data, with a flexible interface for defining regions of interest by administrative boundaries, protected areas, or custom polygons. Validation demonstrated high accuracy, with over 97% agreement with published reference data. This application makes complex geospatial analysis accessible to the mass people, including researchers, land managers, policymakers, and general people, empowering them to support conservation and environmental policy without requiring extensive technical expertise. Our ongoing work is focused on expanding the platform to include other critical environmental indices and a dedicated module for land cover change and transition analysis, further enhancing its utility for monitoring the health of global ecosystems.

## Ethics statements

### CRediT author statement
**Md. Moktader Moula:** Conceptualization, Writing- Original draft preparation, Methodology, Validation, Formal Analysis, Data curation. **Israt Jahan Shonom**: Writing- Original draft preparation, Visualization. **Azharul Islam:** Writing- Reviewing and Editing, Investigation. **Mohammad Mosharraf Hossain:** Writing- Reviewing and Editing, Validation, Investigation

### Acknowledgments
This research did not receive any specific grant from funding agencies in the public, commercial, or not-for-profit sectors.

### Declaration of interests
The authors declare that they have no known competing financial interests or personal relationships that could have appeared to influence the work reported in this paper.